# Face-to-face anneal temperature controls lattice parameter in Ta(C,N) virtual substrates for AlGaN power electronics

Noah Zahn[1,2†], Julia L. Martin[2†], Michelle A. Smeaton[2], Renae Gannon[2], Henry Garland[2], and M. Brooks Tellekamp[2*]

[1]Rossin College of Engineering, Lehigh University, Bethlehem PA 18014, United States

[2]National Laboratory of the Rockies, Golden CO 80401, United States

**ABSTRACT:** Tantalum carbide (TaC) thin film 'virtual' substrates are a highly desirable material for $Al_{0.5}Ga_{0.5}N$ vertical power electronics devices due to lattice matching, thermal expansion matching, and metallic conductivity. However, the material has not been demonstrated to support $Al_xGa_{1-x}N$ layers of variable composition *x*, limiting the range of device applications. We present a method to achieve tunable rock salt $TaC_xN_{1-x}$ virtual substrates via a face-to-face annealing of TaC thin films in a $N_2$ atmosphere. The results suggest that annealing temperatures below 1600 °C promote partial uptake of nitrogen onto carbon and anion vacancy sites to form rock salt Ta(C,N) with intermediate anion compositions. At temperatures ≥ 1600 °C, nitrogen primarily occupies the anion sublattice and the crystalline quality and surface morphology simultaneously degrade coincident with the formation of secondary phases. This study demonstrates the growth and processing parameters necessary to make tunable lattice constant virtual substrates for $Al_xGa_{1-x}N$ from x = 0.5 – 1, enabling vertically conducting power electronic devices with reduced defect density at high Al-content.

## 1. INTRODUCTION

Thermal processing is important across semiconductor science to improve crystal quality and properties, heal defect damage, and promote efficient carrier transport between metals and semiconductors. Annealing to activate Mg-dopants in p-type GaN by driving out compensating hydrogen was a critical technological achievement leading to the widespread adoption of light emitting diodes (LEDs) for solid state lighting applications. High-temperature annealing (> 1500 °C) has emerged as a critical step for III-N UV-LEDs and power devices to reduce defect densities in AlN virtual substrates (also known as templates). In this process, sputter-deposited AlN is annealed at high temperatures in a face-to-face geometry to reduce extended defects. These virtual substrates replace the previous chemical vapor deposited AlN templates, which are often > 4 µm thick, with a layer only 400 nm thick. AlN is however electrically insulating, requiring a semiconducting buffer layer in a pseudo-vertical geometry for vertical conducting applications such as high-currents needed for next-generation power devices.

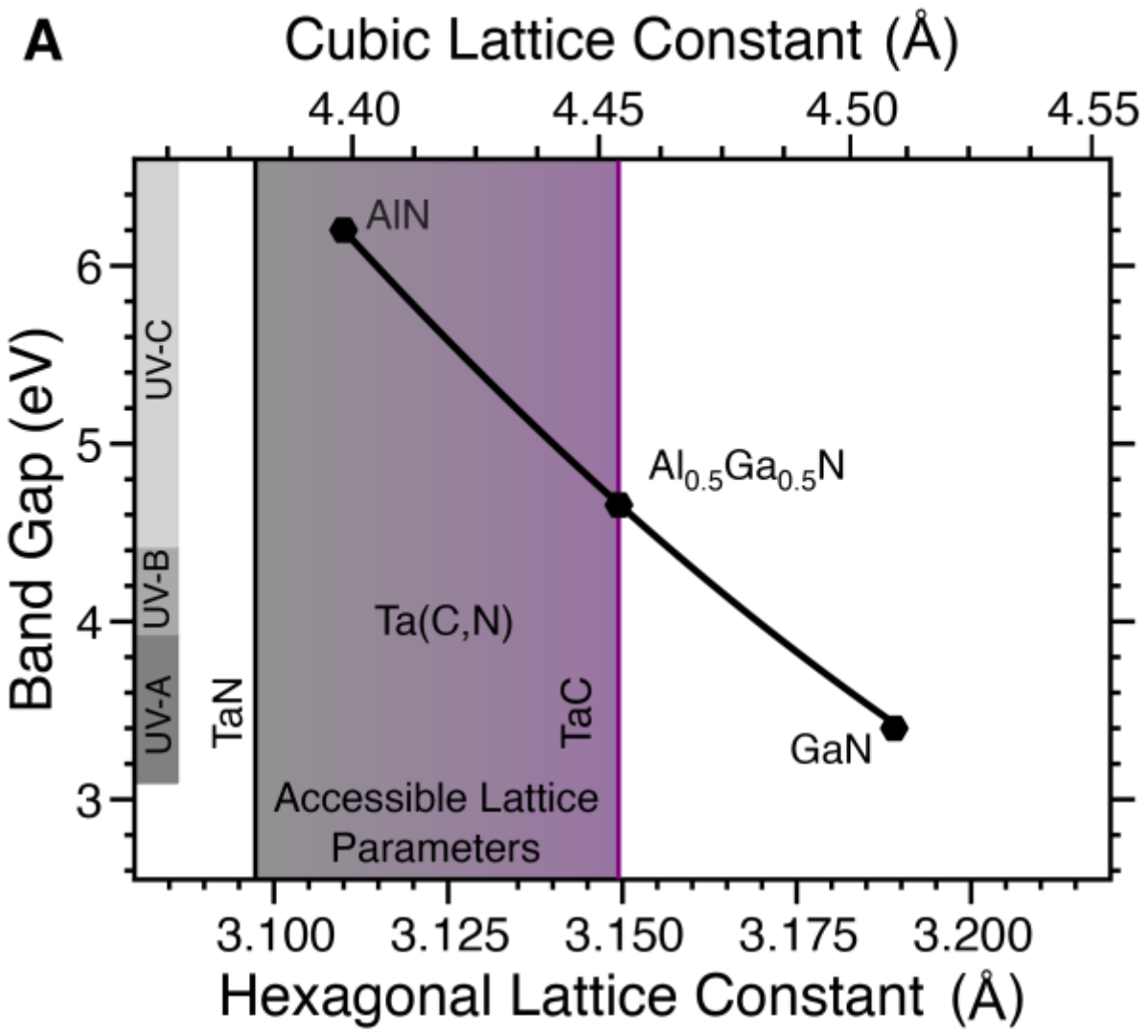

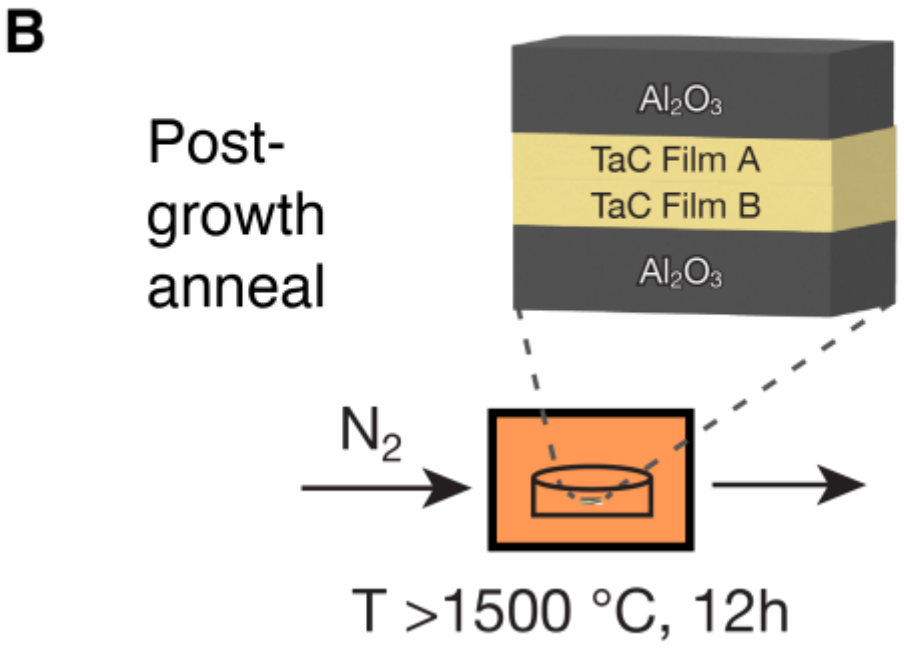

*Figure 1: A) Band gap versus hexagonal lattice constant space for the AlN-GaN alloy system. The TaC and TaN (111) pseudo-hexagonal coincident lattice parameter is overlayed on the hexagonal axis with the true cubic lattice parameter on the top x-axis. The (111) lattice parameter of Ta(C,N) alloys is lattice-matched to $Al_xGa_{1-x}N$ from x=1 – 0.5. B) Illustration of the face-to-face nitrogen annealing geometry in which two separate TaC films on sapphire substrates are placed face-to-face within a tray inside the furnace.*

As global energy demands increase, electrification of industrial, transportation, and consumer sectors becomes crucial.[1] Silicon-based semiconductors have been the selected materials family for decades, yet a reasonably narrow bandgap restricts operating temperature and voltage.[2] Transitioning to ultrawide-bandgap (UWBG) materials ($E_g$ > 3.4eV) carries numerous advantages including a high critical electric field, lower on-state resistances, and higher frequency operation which will decrease the size of passive components. Manufacturers utilizing these properties can fabricate, for instance, compact power diodes with excellent efficiency and reliability.[2–4]

$Al_xGa_{1-x}N$ is an attractive candidate material for next-generation power electronics devices due to a tunable direct bandgap (~3.4 – 6.2 eV) that spans the ultraviolet bands (Figure 1A), high carrier mobility, high breakdown field, bipolar dopability, and mature synthesis. $Al_xGa_{1-x}N$ is n-type dopable for x ≲ 0.85, making compositions between x=0.5 and x=0.85 most desirable for power applications due to the combination of dopability and band gap energy.[5,6] Widespread implementation of $Al_xGa_{1-x}N$ is however restricted due to the lack of substrates suitable for epitaxial growth, and the insulating nature of current substrates. Bulk AlN substrates have been explored, however they are not lattice-matched to the dopable intermediate compositions, and are not conductive for vertical device architectures.[7] Recently, (111)-oriented TaC was demonstrated as a conductive and lattice-matched substrate for AlGaN epitaxy where the (111) hexagonal close-packed plane of TaC is commensurate with the basal plane of ~$Al_{0.5}Ga_{0.5}N$.[8] The study showed that high-temperature, face-to-face annealing of these substrates generated a step-and-terrace grain structure, and that heteroepitaxy of AlGaN on the TaC virtual substrates demonstrated interfaces absent of misfit dislocations. The cohesiveness of the structure supports the usage of TaC but was only demonstrated for a single $Al_{0.7}Ga_{0.3}N$ composition.[8] Conductive virtual substrates designed for a range of compositions x in $Al_xGa_{1-x}N$ could significantly expand design space while improving performance by eliminating strain-based defects (Figure 1A)

In this work, we study the structural, chemical, and morphological effects of face-to-face annealing temperature on epitaxial (111)-oriented TaC thin films grown on sapphire substrates by RF sputtering. As illustrated in Figure 1B, we anneal these thin films in nitrogen environments at temperatures ≥ 1525 °C to improve surface morphology and drive a gas-solid reaction between TaC and $N_2$. Alloying in the TaC-TaN system is explored to control the lattice parameter and engineer lattice-matched virtual substrates for $x = 0.5 – 1$ in UWBG $Al_xGa_{1-x}N$. The lattice-matched space between Ta(C,N) and $Al_xGa_{1-x}N$ is shown in Figure 1A. We demonstrate variations in crystalline phase, composition, lattice parameter, surface morphology, and grain tilt of the films with annealing temperature, placing practical limits on annealing temperatures for these virtual substrates. We hypothesize that nitrogen occupies carbon vacancies in the $TaC_{1-x}$ at lower anneal temperatures, and substitutes for carbon at higher anneal temperatures. This method demonstrates virtual substrates lattice-matched to $Al_xGa_{1-x}N$ across a wide and technologically relevant composition range for next-generation power electronics devices.

## 2. METHODS

*Thin Film Deposition*

(111)-oriented TaC thin films were prepared via radio-frequency (RF) sputtering of a 3" compound TaC target (Princeton Scientific) onto multiple co-loaded 10x10mm (0001)-oriented sapphire substrates. The substrates were cleaned via solvent and piranha solution to remove organic contaminants. The back side of each substrate was metalized with Ta prior to sputtering to promote infrared absorption from the radiative substrate heater. Films were deposited for 120 minutes at 800°C with a power density of 0.02 W/mm$^2$ (100W applied) in 2 mTorr and 20 sccm Ar. The sample holder was rotated for uniformity among the co-loaded substrates.

*Face-to-Face Annealing*

Films were annealed to drive the carbide-to-nitride thermal conversion using an 1800 Series CM Furnace. Samples were annealed in pairs in a face-to-face setup to prevent decomposition of the films and improve thin-film crystallinity (Figure 1B). Films were placed into 11x11 mm recesses on an alumina anneal tray which was covered by an identical tray to minimize thermal gradients between the upper and lower samples. The trays containing the samples were then inserted into the center of the tube furnace, which was held at a base temperature of 300 °C. A 5–10 minute nitrogen gas purge was then initiated at approximately 25 sccm to remove oxygen from the environment. The nitrogen hose remained open throughout the heat treatment to ensure oxygen presence was minimal. The partial pressure of oxygen, $pO_2$, was monitored in the exhaust stream and was $< 10$ ppm before increasing the temperature. The furnace was ramped to 1200 °C at 5 °C/minute, and above this temperature at 1 °C/minute. Samples were then reacted at 1525 – 1650 °C for 12 hours. The cooling and heating ramp profiles were symmetric. $pO_2$ was occasionally evaluated during each of the heat treatments, and it was observed to increase with elevated annealing temperature. The maximum $pO_2$ value observed was ~25 ppm during the 1650 °C anneal.

*Thin-Film Characterization*

A Rigaku SmartLab diffractometer equipped with a Cu-Kα source was used for XRD measurements. The X-rays were monochromated with a 2-bounce Ge channel-cut crystal and measured using a proportional counting detector. Rocking curve measurements were obtained around the (111) and (113) reflections in symmetric and asymmetric geometries, respectively, by rocking the incident angle (ω) and holding azimuth (χ) and rotation (ϕ) constant. Rocking curve peak fitting was performed by least squares minimization using a Pearson-VII profile.

A Bruker Dimension Icon instrument was used to evaluate surface morphology using Bruker AFM Tespa-V2 tips. 10 x 10 μm and 1 x 1 μm scans were collected near the center of each film at a scan speed of 0.4 – 0.5 Hz and a resolution of 512 pixels. Gwyddion software was used for processing to flatten and record RMS surface roughness, average facet angle, and facet angle distribution.

Compositional analysis was performed by wavelength-dispersive X-ray fluorescence (WD-XRF) on a Rigaku ZSX Primus IV. Samples were irradiated with X-ray emission from a Rh tube with applied voltage and current of 50 kV and 50 mA, respectively. The instrument is equipped with a range of diffractometer crystals for wavelength dispersion ranging from 200-2.848 Å to probe B to Cm, respectively. Due to the light mass of nitrogen, a precise composition quantification with WD-XRF using Rigaku internal standards is difficult. As such, quantitative elemental analysis was performed via electron probe microanalysis (EPMA) at the University of Oregon CAMCOR shared facilities.

To visualize the film structure and composition, a thin specimen for Scanning Transmission Electron Microscopy (STEM) was prepared using Xenon Plasma Focused Ion Beam (PFIB) and Scanning Electron Microscopy (SEM) on a Thermo Fisher Scientific Helios 5 PFIB CXe using standard PFIB lamella preparation techniques[7]. STEM imaging was performed on a Thermo Fisher Scientific Spectra 200 STEM operated at a 200 kV accelerating voltage with a 24.2 mrad convergence semi-angle. STEM electron energy loss spectroscopy data were acquired using a Gatan Enfinium spectrometer in dual EELS mode to capture the C-K, N-K, O-K, Al-K, and Ta-M edges simultaneously.

## 3. RESULTS

### 3.1. Structural and Composition

Structural changes to TaC thin films during high-temperature face-to-face annealing were analyzed by XRD performed on as-grown and post-annealed films. The as-grown TaC thin films are primarily (111)-oriented and rotationally twinned in-plane due to the epitaxial relationship with the c-plane sapphire substrate. There is also a small fraction of (200) cubic TaC observed by XRD.[6] The structural changes are shown in Figure 2 by symmetric 2θ-ω XRD in order of ascending annealing temperature from bottom to top. The as-grown film is displayed at the bottom of the figure. The vertical dashed lines designate indexed peaks for various potential phases containing Ta, N, and O including hexagonal TaN (orange), rock salt TaN (purple), and TaON (gray). The target rock salt TaC (111) and TaN (111) peaks, which occur at 34.75° and 35.12°, respectively, are shown as a 2θ range with a gray-to-purple gradient. Reacting the films at 1600 °C or greater results in the strong formation of TaON, hexagonal TaN, and the cubic (200)-orientation of rock salt TaN. At temperatures above 1550 °C we observe a shift in the (111) reflection to higher angles, consistent with the uptake of nitrogen on anion sites that shrink the lattice. Figure 2 also illustrates the formation of hexagonal TaN at temperatures as low as 1525 °C.

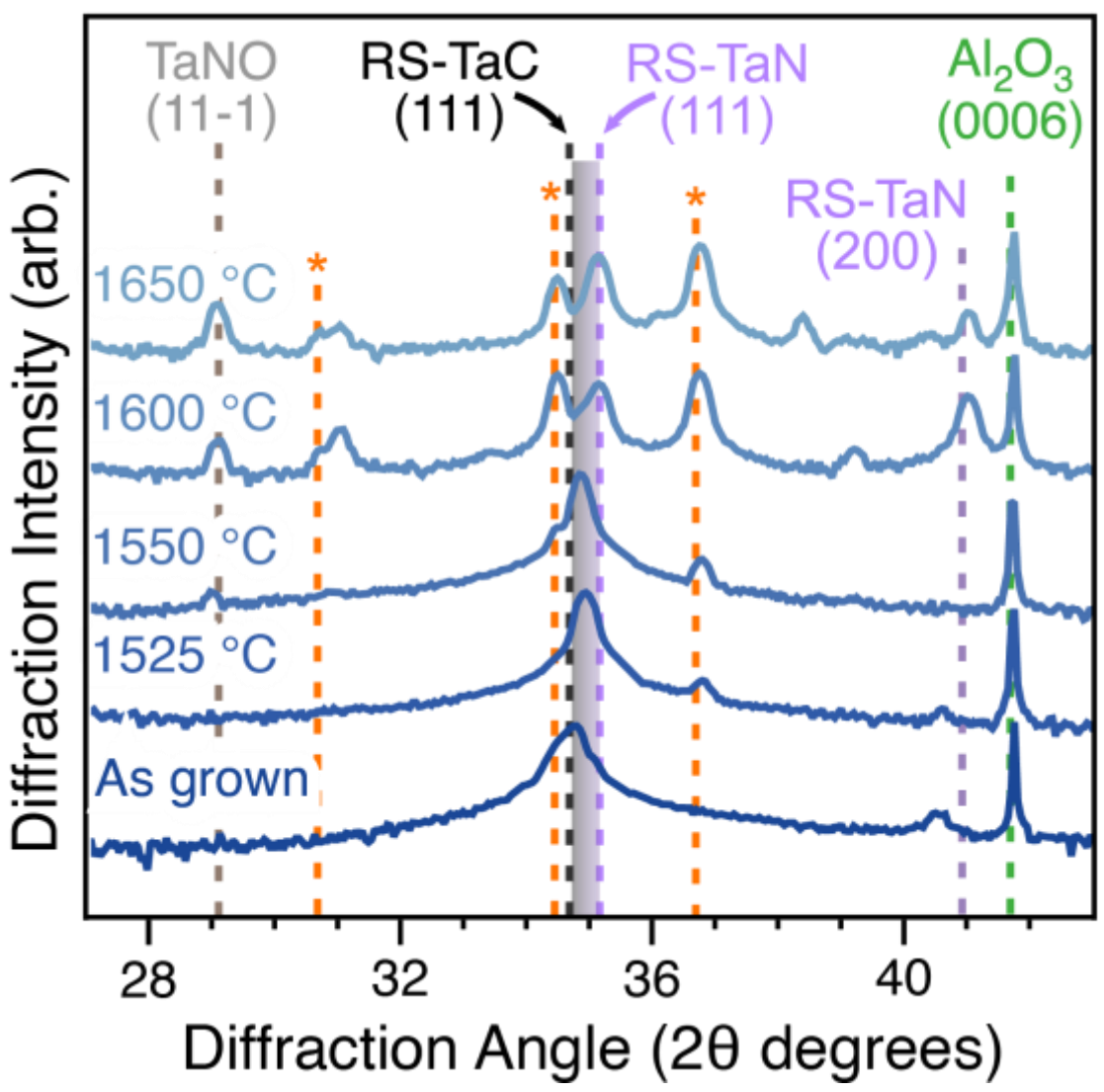

*Figure 2: Indexed XRD spectra comparison for TaC thin films in as grown and post-annealed conditions. Wide-angle scans were completed on samples annealed at 1525°C, 1550°C, 1600°C, and 1650°C, as well as a pre-annealed sample. As anneal temperature increases the (111) rock salt peak shifts from the TaC (111) position to the TaN (111). Labels stating the annealing condition are given along the right axis. Orange dashed lines denote the peaks corresponding to the hexagonal phase of TaN, which was also witnessed in numerous films.*

The d-spacing values of the (111) planes were calculated from the Bragg condition for diffraction using the angle of highest intensity in each peak from high-resolution 2θ-ω scans. These values were compared with anneal temperature in Figure 3A. Theoretical d-spacing values of the (111) TaC and (111) rock salt TaN planes are labeled as horizontal dashed lines, with a secondary y-axis highlighting the equivalent $Al_xGa_{1-x}N$ compositions which would align with bulk strain-free (111)-oriented TaC. The epitaxial alignment used for this assignment is TaC (220) || AlGaN (110).[8] The (111) d-spacing value for the as-grown film was ~0.009 Å greater than pure TaC. This observation can be attributed to crystal strain created by thermal expansion or lattice mismatch, as described by prior works.[8] Films annealed at 1600-1650°C have the lowest d-spacings while those annealed at intermediate temperatures have intermediate lattice spacing that roughly compares to that of theoretical rs-TaC, with ~0.01-0.02 Å variation above or below it.

To understand the relationship between annealing temperature, vacancy concentration, anion composition, and epitaxial strain, quantitative compositional analysis was collected via EPMA and compared with annealing temperature in Figure 3B. This data in combination with Table S1 confirms the idea that as-grown films are slightly carbon deficient, while films annealed at 1525-1550°C contain mixed nitrogen and carbon on the anion site of the rock salt structure and approach stoichiometric in terms of composition. Notably, Figure 3A demonstrates that the annealed films have intermediate lattice spacing between TaC and TaN and contain both nitrogen and carbon. The nitrogen to carbon ratio is greater for higher annealing temperatures, and the anion concentration of film annealed at 1650 °C is ~75% nitrogen. The shift of lattice constant and increase in nitrogen signal relative to carbon indicates a carbonitride Ta(C,N) has been formed. To further support the compositional trends revealed by Figure 3B, we qualitatively explored the composition of the annealed films via wavelength-dispersive X-ray fluorescence (WD-XRF) which is contained in Figure S1 which confirms increased nitrogen and decreased carbon for increasing anneal temperatures.  Importantly, the Ta signal in Figure S1D remains consistent as a function of anneal temperature, confirming that no significant oxidation of TaC and subsequent $Ta_2O_5$ evaporation has occurred.

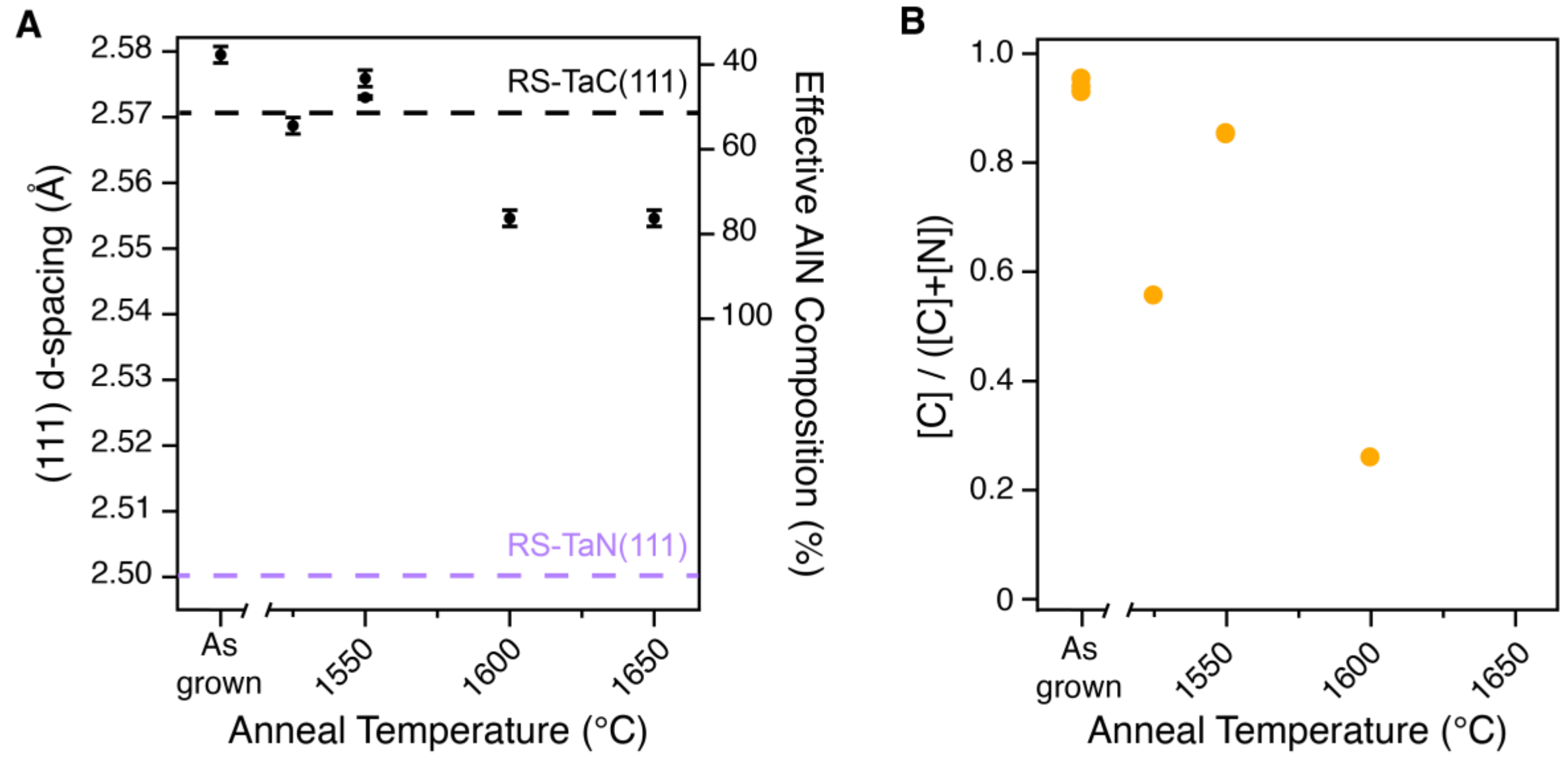

*Figure 3: A) Structural and B) compositional analysis of annealed TaC films. Black points in A) highlight the (111) plane d-spacing as annealing temperature was altered. Error bars were included due to limitations in the step size of the XRD instrument. Dashed lines are drawn showing the theoretical (111) d-spacings of TaC and rock salt TaN. The yelllow points in B) illustrate the corresponding composition of carbon within the film. Increased annealing temperatures coincided with reduced carbon composition.*

Rocking curve diffraction scans of the (111) and (113) reflections of the Ta(C,N) films were collected in symmetric and asymmetric geometry, respectively, to understand changes to crystalline mosaicity as a function of anneal temperature. The as-grown sample full-width at half maximum (FWHM) of the ω rocking curve is compared with annealed samples in Figure 4 as a function of temperature. To account for any sample-to-sample variation, each wafer was measured before and after annealing, and the as-grown FWHM is relatively consistent from sample to sample. For these (111)-textured films, the (111) out-of-plane direction is only sensitive to mosaic tilt, while the (113) reflection contains both in-plane and out-of-plane components and is therefore sensitive to both mosaic tilt and twist.[8,9] A two-phase structure was observed in each of the as-grown (111) peaks – this has been previously reported both for sputtered TaC and other materials such as AlN and is related to a grain tilting relaxation mechanism after the film exceeds a critical thickness.[6,10] The broad and narrow components of these rocking curves were fit with two curves for (111) pre-anneal films only. After annealing, the rocking curves increase in intensity and merge from a two-phase structure to a single peak.

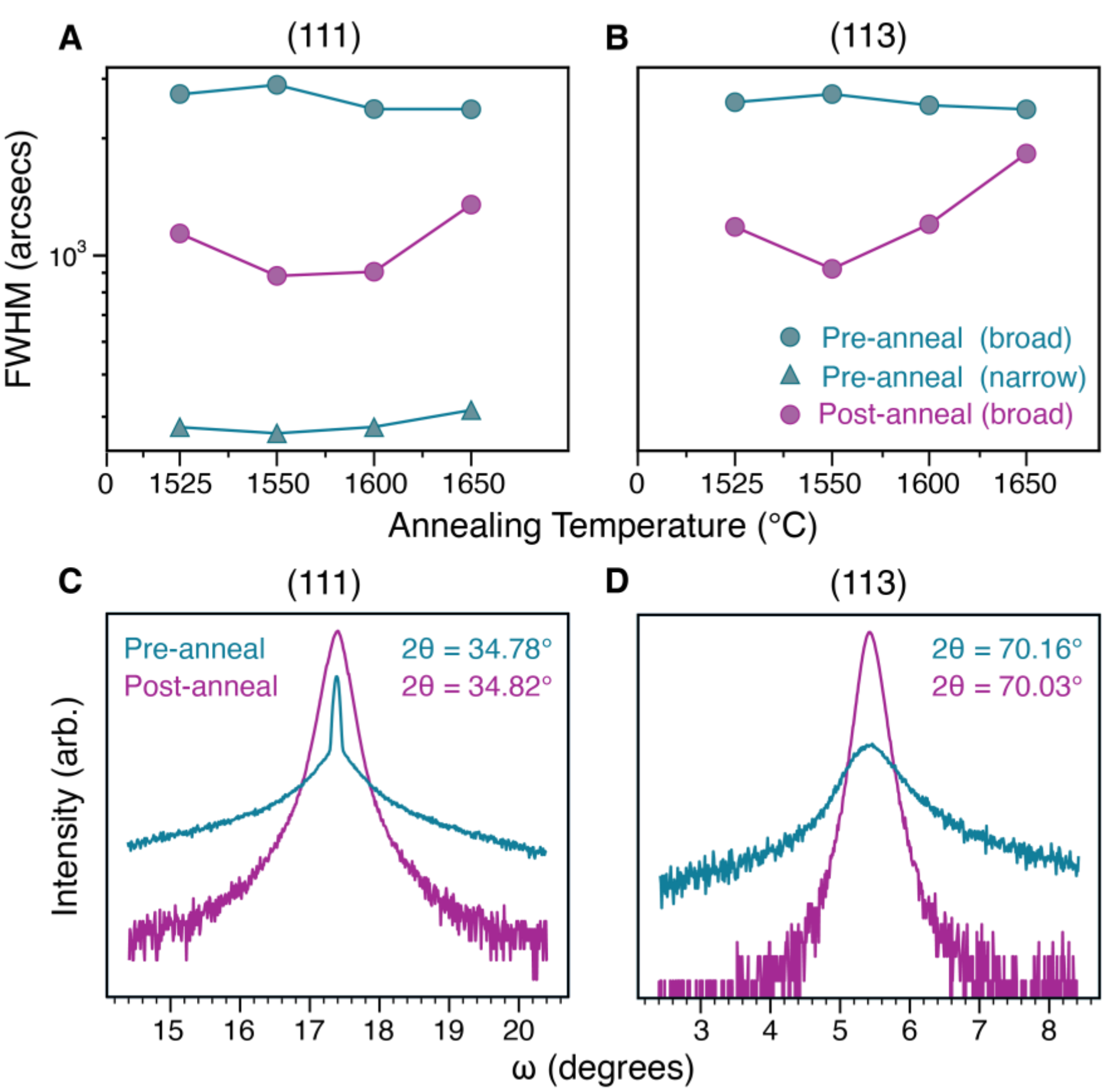

*Figure 4: Rocking curve FWHM data for TaC thin films in as grown and post-annealed states. Symmetric (111) and asymmetric (113) reflections were scanned (A and B, respectively) to illustrate variations in crystalline quality with annealing temperature. Representative pre- and post-anneal rocking curve plots are shown to demonstrate reductions in FWHM following annealing for typical (111) and (113) scans in C and D, respectively.*

The FWHM of the broad peak decreased following annealing for both the (111) and (113) reflections, indicating an improvement in crystalline quality. The reflection intensity also increases while the background intensity decreases, indicating that a majority of highly mistitled grains are reoriented during the face-to-face anneal. The post-anneal crystalline quality degraded with increasing temperature; beyond 1550 °C the FWHM of both peaks tended to increase relative to the 1550 °C point, and the lowest FWHM for both the (111) and (113) reflections occurred at an annealing temperature of 1550 °C. Relative to the as-grown conditions, the crystalline quality was improved in the annealed samples regardless of annealing temperature. Studies at still higher temperatures could demonstrate degradation of the films beyond their as-grown states.

### 3.2. Surface Morphology

AFM was used to measure the post-anneal surface morphology of each film and demonstrated significant changes as a function of annealing temperature, as shown in Figure 5. Step-and-terrace structures are clearly visible in Figure 5A at 1525 °C, the lowest annealing temperature studied. Terrace size increased along with annealing temperature, with the terraces in Figure 5G for a film annealed at 1650 °C appearing the largest across each of the 10 $\mu m^2$ scans. RMS roughness values collected from the 10 µm scans of each film demonstrate a similar trend. The lowest roughness was measured at 5.2 nm RMS for an anneal temperature of 1550°C before eventually increasing to 53 nm RMS for a film annealed at 1650°C.

Abrupt and well-defined terrace edges are visible for films annealed at 1525°C and 1550°C in Figure 5B and Figure 5D, respectively. Beyond 1550°C, the terraces become less distinct. The terraces become poorly defined at 1600 °C, demonstrating significant height variation in Figure 5F. Streaks and blurs were seen in Figure 5H for the film annealed at 1650°C in what appears to be stacked terraces along the edge of grains. The clearest terraces occurred for the sample annealed at 1550°C, where sharp variations in film height were visible surrounding all terraces. Thin, sharp walls of material appeared in all the 1 $\mu m^2$ scans as bright white streaks. The origin of these features is unknown but likely related to strain and relaxation in the extreme temperature annealing process. We note that other growth and annealing conditions do not contain these domain walls, and further work is ongoing to understand and control these features.

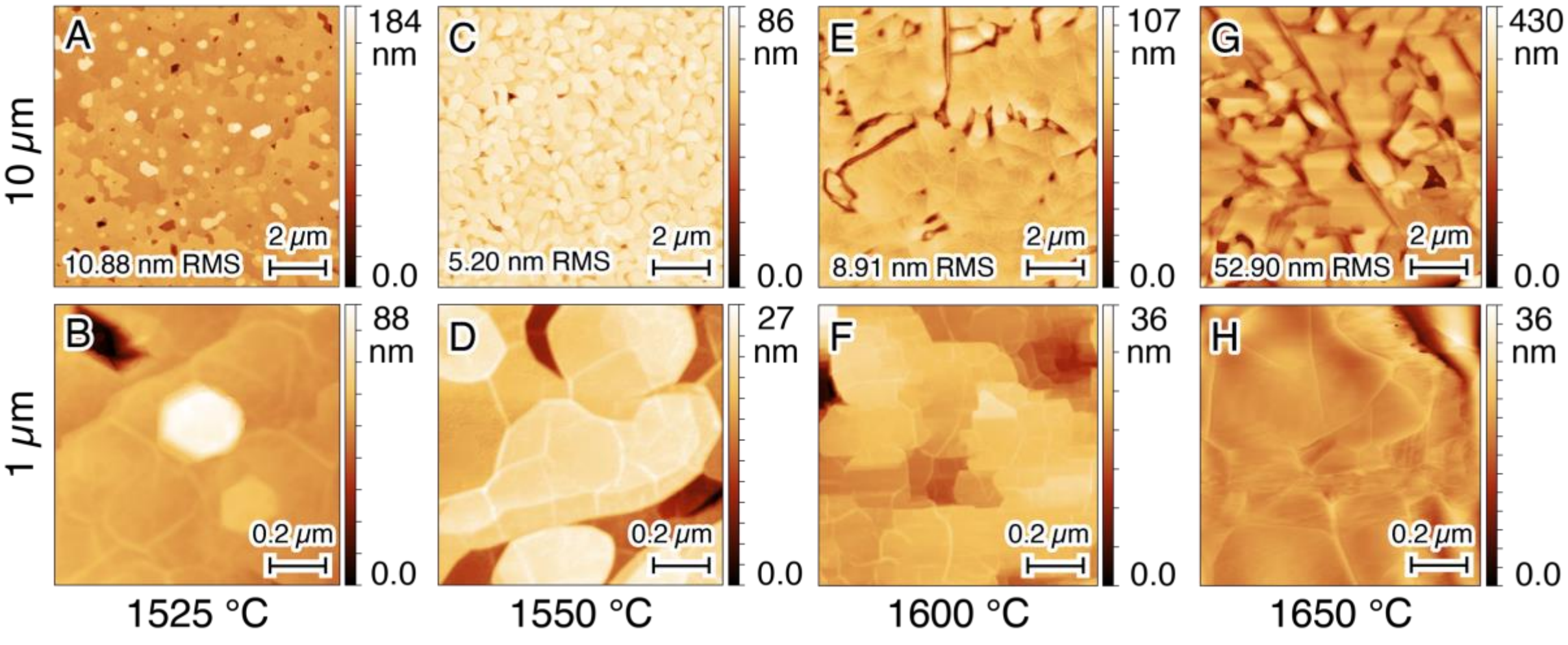

*Figure 5: AFM plots illustrating the surface morphology of each TaC film. 10µm and 1µm scans were taken of films annealed at 1525 ℃ (a and b), 1550 ℃ (c and d), 1600 ℃ (e and f) and 1650 ℃ (g and h). An intensity bar adjacent to each micrograph shows the z-height corresponding to the variety of shades present.*

Additional analysis showing the distribution of facet tilt angles can be found in Figure S3. Films annealed at low temperatures carried a tight spread with facets tending to be oriented in one of 6 evenly spaced angles. Those annealed at high temperatures carried a much broader distribution without containing any preferential facet tilt angles, corresponding to the loss of faceting seen in the Z-height images above.

## 4. DISCUSSION

In previous work, we demonstrated (111)-oriented TaC as a conductive and lattice-matched substrate for AlGaN epitaxy, and the resulting MBE-grown AlGaN film was high in Al content (~70%), strain free, and demonstrated an abrupt interface with the TaC layer. This demonstration, combined with recent computational studies[12] motivate have motivated exploration of rock salt Ta(C,N) alloys that provide controlled the lattice parameter to expand the range of feasible AlGaN compositions. Among synthetic approaches, the formation of mixed-anion carbonitrides has already been observed in high temperature $N_2$ annealing for $TiC_xN_y$, $HfC_xN_y$, and (W, Mo) (C, N) powders.[15–17] As such, we chose to explore carbide-to-nitride conversion of sputter-grown TaC films via high temperature thermal annealing in a nitrogen environment. We employed temperatures above 1500 °C to enable nitrogen reactivity and diffusion into the solid, while also promoting improved crystallinity.[11–13]

XRD studies of the annealed films probing structural changes demonstrate a shift of the rock salt (111) peak which, coincident with the EPMA-measured nitrogen content in Figure 3B, supports formation of a mixed anion Ta(C,N) rock salt phase. Further, the XRD studies ultimately revealed two distinct temperature regimes within the range studied herein (1525-1650 °C): "intermediate" temperatures at or below 1550 °C and "high" temperatures above 1550 °C. Films annealed at 1525 – 1550 °C show intermediate nitrogen anion compositions ranging from $x = 0.14$ to $x = 0.44$ in $TaC_xN_{1-x}$, corroborated by both structural and compositional measurements. The mixed phases at these temperatures carry common hexagonal lattice parameters corresponding to AlN compositions in $Al_xGa_{1-x}N$ of $x = 0.43$ - $0.54$ ignoring strain. As previously described, biaxial strain created during film deposition results in films with compressively strained out-of-plane lattice parameters, while the in-plane lattice parameters are larger corresponding to greater x in $Al_xGa_{1-x}N$ than would be assumed from the bulk cubic lattice. Indeed, we find that the range of lattice-matching for this film set is closer to $x = 0.6 – 0.7$ for the 1525 – 1550 °C anneals. Minimizing strain during the sputtering process could reduce the lattice parameter such that films annealed at intermediate temperatures correspond entirely to AlN compositions within the $0.5 < x < 1.0$ range. At these intermediate temperatures, even after a 12-hour reaction, there is not enough thermal energy to remove carbon from the lattice stabilizing the desired rock salt phase.

Additionally, formation of a rock salt Ta(C,N) structure supports literature hypotheses that the rock salt phase is stable at carbon anion compositions greater than x ≈ 0.25.[12]

Samples annealed at high temperatures (1600 °C or greater) are nitrogen-rich in anion content (Figure 3B, EPMA was not obtained for the 1650 °C sample), indicating removal of carbon from the anion lattice. The effective hexagonal lattice parameter for these films corresponds to an $Al_xGa_{1-x}N$ composition x = 0.76. Between 1600 °C and 1650 °C, the rock salt lattice constants do not change, and we therefore conclude that nitrogen substitution on carbon sites is limited to ~25% carbon at practically accessible temperatures, noting the onset of plastic deformation in sapphire at 1700 °C and above.

The reduced lattice constant at anneal temperatures of 1600 – 1650 °C, coupled with the large nitrogen to carbon ratio, is interesting because the rock salt structure in $TaC_xN_{1-x}$ is thermodynamically dominant for $0.25 < x < 1$, however phase separation is likely for $x < 0.5$.[12] While rock salt TaN is metastable, it becomes more stable with increased entropy from vacancies.[18,19] It is hypothesized that the gas-solid exchange reaction from an initial rock salt precursor allowed the structure to remain primarily in the rock salt phase at greater nitrogen compositions,[20] however these results qualitatively support the previously predicted stability window of Ta(C,N).[14] More analysis is needed to quantify the vacancy concentrations and further explore the interplay between composition and phase stability. Annealing at ≥ 1600 °C also results in the formation of multiple Ta-N phases that can be understood by the complex Ta-N phase diagram;[21] at these temperatures a completely homogenous rock salt film was not achieved, with a considerable portion of the TaN forming in the hexagonal phase as seen in Figure 2.

The two-stage process above and below 1600 °C may be driven by initial carbon vacancies on the anion sublattice, enabling nitrogen uptake without driving the gas-solid reaction between TaC and $N_2$. Carbon vacancies are typically characterized by shifts in lattice parameter, however biaxial strain at the interface of the TaC thin films and $Al_2O_3$ has been previously reported[8] and can explain the deviations of the (111) d-spacing compared to bulk TaC seen in Figure 3A. This biaxial strain can shift the (111) XRD peak position, obscuring the presence of carbon vacancies which shrink the lattice. We hypothesize that at lower reaction temperatures, 1525 °C and possibly below, disassociated nitrogen will first occupy anion vacancy sites. Similar behavior has been observed in oxyhydrides where vacancies promote $N^{3-}/O^{2-}$ anion exchange, enabling full nitridation at a reaction temperature of 800°C.[22]

Latent oxygen may also be responsible for nitrogen substitution. The alumina furnace tube, boat, and $Al_2O_3$ substrate will effuse oxygen above ~1600 °C. With an oxygen partial pressure $pO_2$ increasing to ~ 20 ppm during the high temperature soak, latent oxygen can react with carbon and desorb as CO. The additional carbon vacancies created by this reaction are then filled with nitrogen.

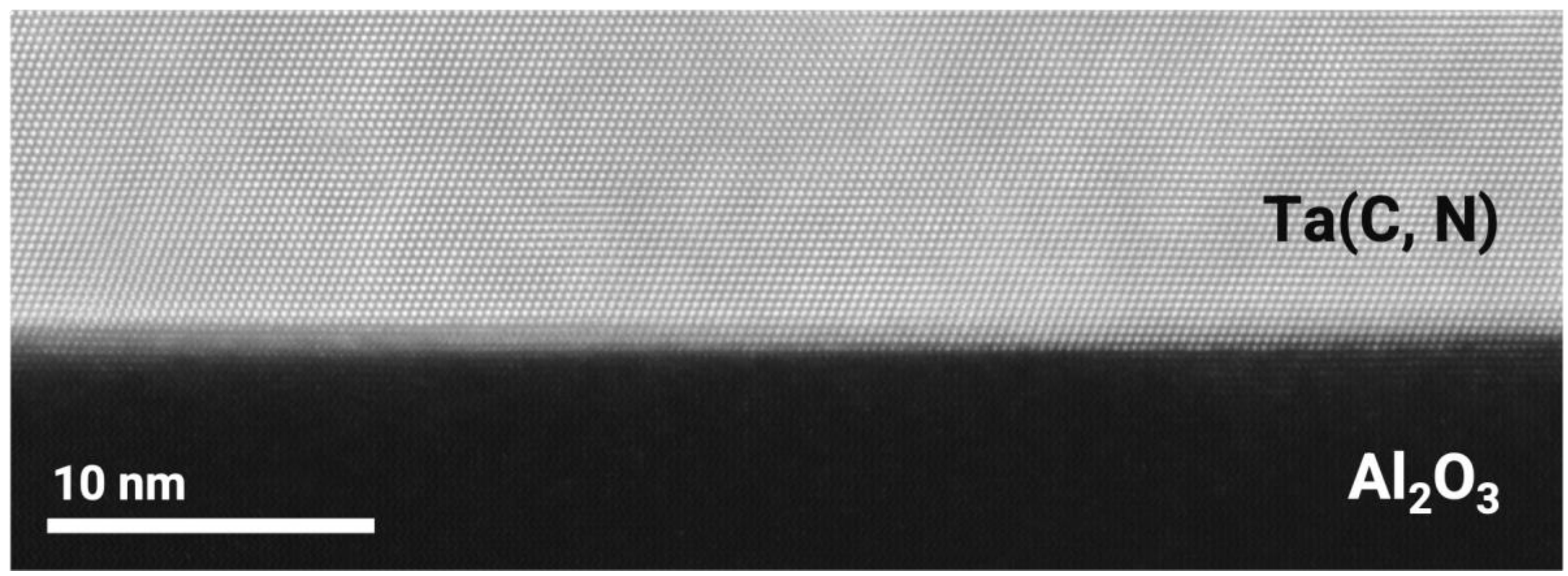

*Figure 6: STEM image of the boundary between a TaC film following a 1550°C nitrogen anneal (upper half) and the underlying $Al_2O_3$ substrate (lower half). While the interface is fairly abrupt, it also exhibits signs of displaced atomic planes as indicated by blurred features at the boundary, supporting the assertion that biaxial strain is present within films. The annealed film has faced partial nitridation and is a mixed Ta(C,N) phase yet still possesses a profile indicative of the rock salt structure, supporting findings from XRD.*

Crystallographic and morphological analysis of the films following annealing continued to support the idea of a broad change in the makeup and structure of the films above 1550°C. Figure 4 and Figure 5 each showed a degradation in the crystalline and structural quality of the films when annealing past this temperature, likely associated with the formation of secondary Ta-N and TaON phases. Mosaicity and facet tilt were indicated to grow more irregular with increasing temperatures by Figure S3, yet improve relative to the as-grown films. Blurring between distinct terraces and overall inconsistent z-heights across the films beginning at Figure 5E-F show the irregularity in facet orientation across the surface. At modest anneal temperatures, the facet distribution on the surface follows the crystallographic directions in a 6-fold pattern, but the facets are no longer oriented at high temperatures (Figure S3). Step-and-terrace surfaces, exposing regular and aligned facets, are important for heteroepitaxial integration of AlGaN on Ta(C,N), and the lower temperature anneals produce a more favorable surface for epitaxial growth.

Further, the structural and chemical uniformity of a film annealed at 1550°C were analyzed using scanning transmission electron microscopy (STEM) which is shown in Figure 6, highlighting the interface between the sapphire substrate and the Ta(C,N) film. STEM imaging confirms the (111)-oriented rock salt structure of the film as well as the presence of some twin domains. Blurring of atomic columns in the image is indicative of strain in the film and no additional phases were observed in the STEM lamella in imaging or EELS (Figure S2). Compositional variation was assessed by EELS in cross-section, and we found that the nitrogen and carbon content in the analyzed area was uniform. The interface was found to be chemically abrupt, and oxygen was not observed in the film. Although hexagonal TaN was seen in XRD, secondary phases were not observed in the 1550 °C annealed film by STEM imaging over multiple investigated locations on the lamella, indicating the phase separation is laterally distributed across the wafer and likely due to radial variation rather than processing conditions.

## 5. CONCLUSIONS

In this work, we demonstrate the formation of rock salt Ta(C,N) through the direct nitridation of TaC thin films by face-to-face annealing at high temperature. The formation of Ta(C,N) with intermediate C:N ratios suggest the possibility of tunable (111)-oriented Ta(C,N) virtual substrates that are lattice-matched to $Al_xGa_{1-x}N$ within a technologically relevant composition range $0.5 < x < 1$. We showed that nitrogen can occupy the anion sublattice and hypothesize that at sufficiently low reaction temperatures, below 1550 °C, the primary mechanism is the uptake of disassociated nitrogen onto anion vacancy sites. Beyond 1600 °C, carbon anions are actively substituted for nitrogen atoms in the gas-solid exchange reaction and a Ta(C,N) film is formed, likely mediated by oxidation of the carbon with alumina as the high temperature oxygen source. The film is primarily rock salt, however higher temperatures promote increasing formation of the ground-state hexagonal TaN and some TaON. This temperature also corresponds to degradation in the overall crystalline and morphological quality of the film, placing an upper bound on processing parameters. For thin film virtual substrates, ideal temperatures of 1550 °C or less are preferred to maintain acceptable crystal quality and surface morphology while avoiding the formation of secondary phases. Overall, this work demonstrates a methodology to create conductive virtual substrates lattice-matched to any $Al_xGa_{1-x}N$ composition between $x = 0.5$ and $x = 1$.

## ASSOCIATED CONTENT

### Supporting Information

The following supporting information is available free of charge:

- WD-XRF qualitative elemental analysis spectra of post-annealed TaCN thin films
- Spectroscopic analysis of a film annealed at 1550°C using EELS and EDS
- Table of composition from EPMA measurements
- Facet tilt angle distribution maps

## AUTHOR INFORMATION

### Corresponding Author

**Brooks M. Tellekamp**—National Laboratory of the Rockies, Golden 80401, United States; Email: brooks.tellekamp@nlr.gov; orcid.org/0000-0003-3535-1831

### Authors

The manuscript was written through contributions of all authors. All authors have given approval to the final version of the manuscript.

## ACKNOWLEDGEMENTS

**Noah Zahn** — Lehigh University, Bethlehem 18015, United States; Email: ntz327@lehigh.edu; orcid.org/0009-0004-8441-2834

**Julia L. Martin** — National Laboratory of the Rockies, Golden 80401, United States; Email: julia.martin@nlr.gov; orcid.org/0000-0001-7609-9922

**Brooks M. Tellekamp**—National Laboratory of the Rockies, Golden 80401, United States; Email: brooks.tellekamp@nlr.gov; orcid.org/0000-0003-3535-1831

**Henry Garland** — Yale University, New Haven 06501, United States; Email: henry.garland@yale.edu; orcid.org/0009-0005-1611-1228

This work was authored in part by the National Laboratory for the Rockies (NLR) for the U.S. Department of Energy (DOE), operated under Contract No. DE-AC36-08GO28308. This work was primarily supported as part of A Center for Power Electronics Materials and Manufacturing Exploration (APEX), an Energy Frontier Research Center funded by the U.S. Department of Energy, Office of Science, Basic Energy Sciences. This work was supported in part by the U.S. Department of Energy, Office of Science, Office of Workforce Development for Teachers and Scientists (WDTS) under the Science Undergraduate Laboratory Internships (SULI) Program. We would also like to acknowledge Julie Chouinard for the EPMA work which made use of the Cameca SX100 EPMA at the University of Oregon CAMCOR shared facilities.

# Supporting information for:

## Face-to-face anneal temperature controls lattice parameter in Ta(C,N) virtual substrates for AlGaN power electronics

Noah Zahn[1,2†], Julia L. Martin[2†], Michelle A. Smeaton[2], Renae Gannon[2], Henry Garland[2], and M. Brooks Tellekamp[2*]

[1]Rossin College of Engineering, Lehigh University, Bethlehem 18014, United States

[2]National Laboratory of the Rockies, Golden 80401, United States

**Corresponding author:** M. Brooks Tellekamp, brooks.tellekamp@nlr.gov

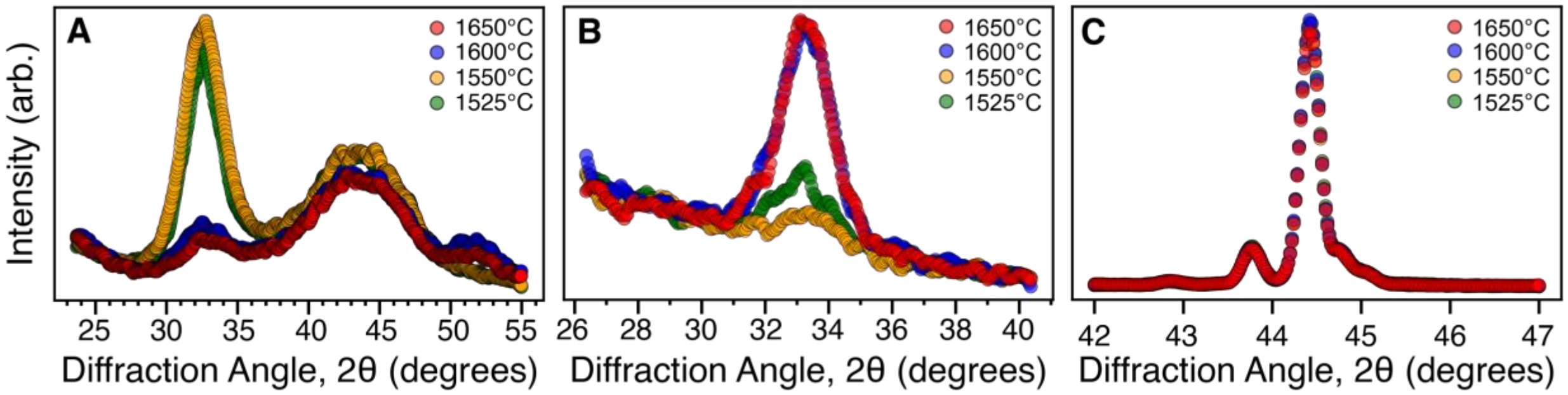

*Figure S1: Wavelength-dispersive X-ray fluorescence (WD-XRF) spectra illustrating qualitative compositional analysis for each TaC film. Scans were collected for A) C-Kα, B) N-Kα, and D) Ta-Lα fluorescence lines for films annealed from 1525°C-1650°C. Increased annealing temperatures correlated with reduced carbon signals and elevated nitrogen signals.*

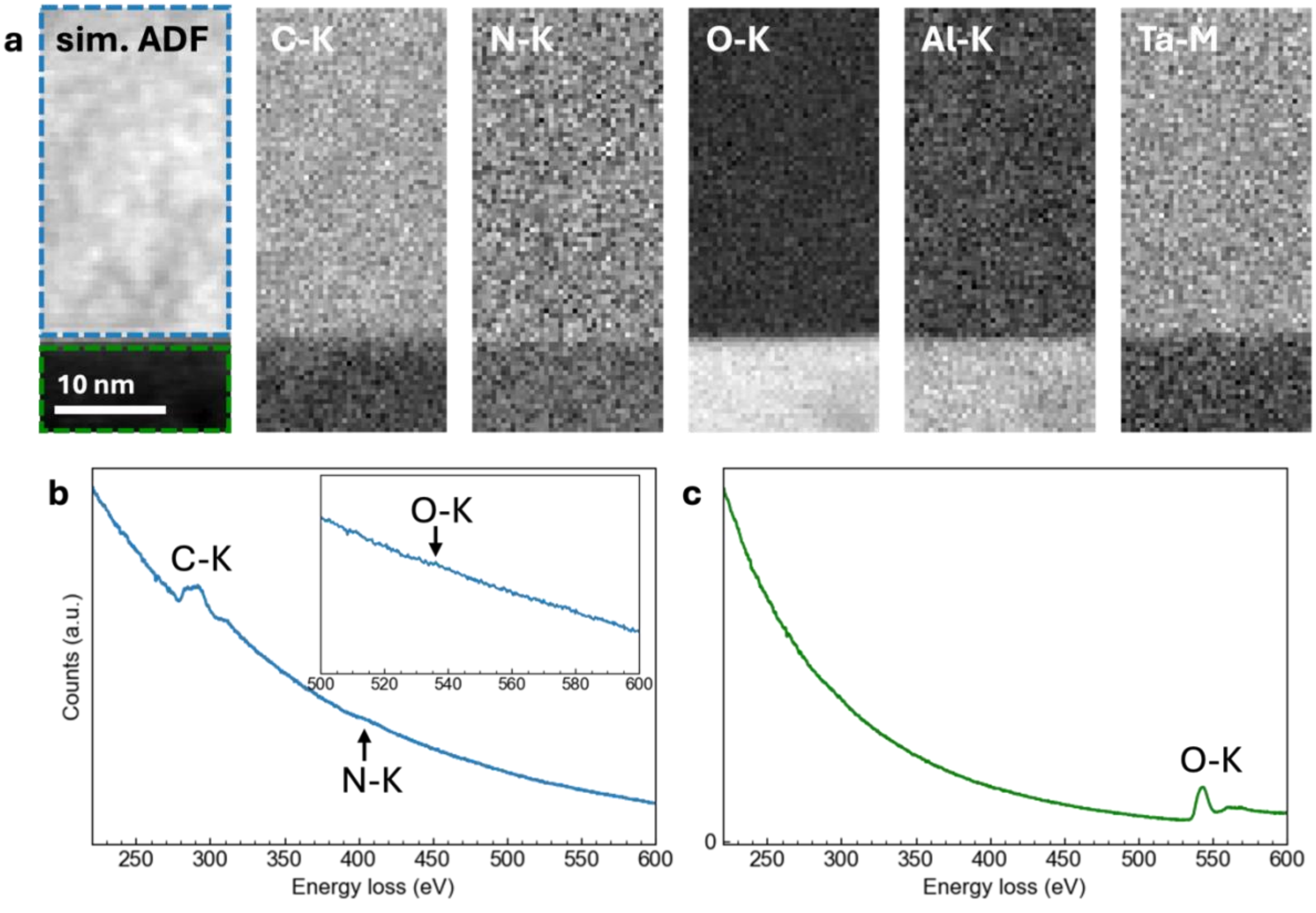

*Fig S2. (a)STEM-EELS map of C-K, N-K, O-K, Al-K, and Ta-M edge intensity across the Ta(C, N)/$Al_2O_3$ interface, showing a sharp delineation in composition and no additional phases. Each map has been normalized individually, meaning intensities are not directly comparable. (b,c) Non-background subtracted EEL spectra summed over the (b) Ta(C, N) and (c) $Al_2O_3$ regions of the map in (a) as indicated by dashed blue and green lines, respectively. The inset in (b) shows a zoomed in region of the spectrum around the O-K edge at ~532 eV, exhibiting a minor bump, which is negligible in comparison to the C, and even low the N signal.*

*Table S1: Composition and anion fractions for three representative as-grown TaC films and several representative Ta(C,N) films at four different annealing temperatures. For anion fractions, only carbon and nitrogen were considered.*

| Sample ID | Anneal Temperature (°C) | $\frac{[\mathrm{C}]}{[\mathrm{C}]+[\mathrm{N}]}$ | $\frac{[\mathrm{anions}]}{[\mathrm{anions}]+[\mathrm{Ta}]}$ |
|---|---|---|---|
| C10_0665_K | 1525 | 0.56 | 0.40 |
| C10_0673_B | 1550 | 0.86 | 0.45 |
| C10_0666_L | 1550 | 0.85 | 0.44 |
| C10_0666_C | 1650 | 0.26 | 0.37 |
| C10_0665_A | As-grown | 0.93 | 0.33 |
| C10_0673_A | As-grown | 0.96 | 0.45 |
| C10_0666_A | As-grown | 0.94 | 0.40 |

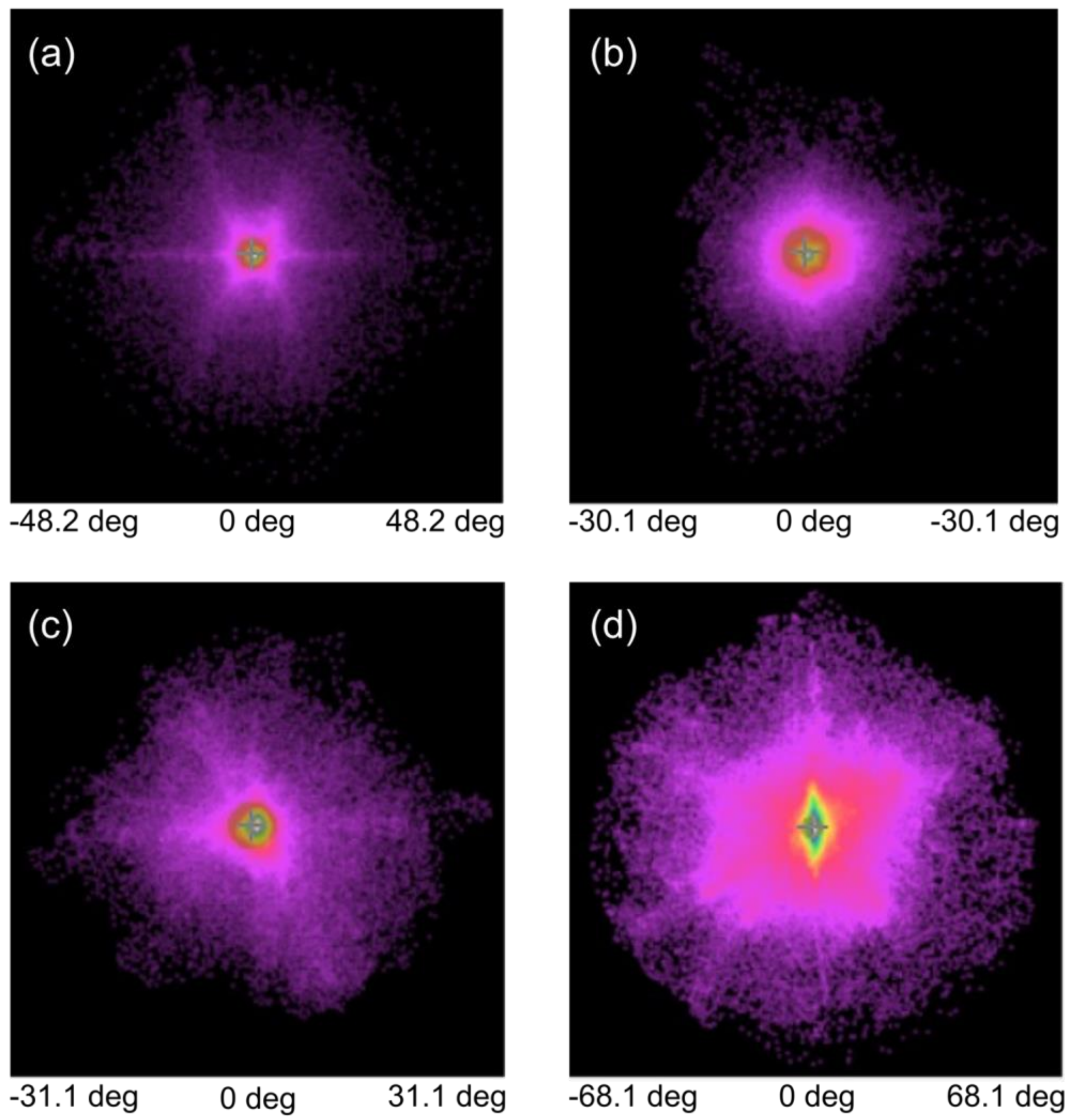

*Figure S3: Facet tilt distribution maps were collected using data from 10μm AFM scans of each TaC film to highlight trends in the regularity of surface facets with annealing temperature. Films annealed at 1525°C (a), 1550°C (b), 1600°C (c), and 1650°C (d) were studied. Horizontal and vertical distances from the center of each plot represent deviation from the majority tilt angle, represented by the blue-green portion surrounding the cursor. Increasing spread in the radius of the pink-shaded region with elevated annealing temperatures represents growing irregularity in the surface morphology of these films.*